\documentclass{mem}
\usepackage{natbib}\usepackage{txfonts}\usepackage{balance}
\usepackage{graphicx}
\idline{75}{282}
\begin{document}
\def\teff{$T\rm_{eff }$}
\def\kms{$\mathrm {km s}^{-1}$}

\title{The low mass end of the IMF}

   \subtitle{}

\author{
C. \,Alves de Oliveira\inst{1} 
}

  \offprints{C. Alves de Oliveira}

\institute{
European Space Astronomy Centre (ESA/ESAC), Science Operations Department, Villanueva de la Ca\~nada (Madrid), Spain
\email{calves@sciops.esa.int}
}

\authorrunning{Alves de Oliveira}

\titlerunning{The low mass end of the IMF}

\abstract{The rapid advances in infrared detector technology over the past decades have impelled the development of wide-field instruments, and shaped our view of the cold universe. Large scale surveys in our Galaxy have discovered hundreds of brown dwarfs enabling the characterisation of the mass function in the substellar regime. I will review the most recent observational results concerning the substellar IMF derived in star forming regions, open clusters, and the field, that must be reproduced and explained by any successful star formation theory.
 }
\maketitle{}

\section{Introduction}

\begin{table*}[ht]
\caption{Selection of results from brown dwarf surveys of the solar neighbourhood.}
\label{fieldimf}
\begin{center}
\begin{tabular}{lllll}
\hline
\\
Survey & Sample & Limit & $\alpha$ & References  \\
\hline
\\
UKIDSS  		&  Late T 			& Mag. limited 		& $-$1$<$$\alpha$$<$$-$0.5 	& Burningham et al. 2013 \\
               		&  mid-L to mid-T  	& Mag. limited 		& $-$1$<$$\alpha$$<$0 			& Day-Jones et al. 2013 \\
WISE  			&  Late T to early Y  	& Vol. limited 		& $-$0.5$<$$\alpha$$<$0 		& Kirkpatrick et al. 2012 \\
CFHTBD  		&  L and T  			& Mag. limited 		& $\alpha$$<$0 					& Reyle et al. 2010 \\
2MASS/SDSS  	&  T  				& Mag. limited 		& $\alpha$$\sim$0 				& Metchev et al. 2008 \\
2MASS		  	&  L  				& Vol. limited 		& $\alpha$$\sim$1.5 			& Cruz et al. 2007 \\
2MASS		  	& MLT 				& Vol./Mag. limited 	& $\alpha$$\sim$0.3$\pm$0.6 	& Allen et al. 2005 \\
\hline
 \multicolumn{5}{c}{In progress} \\
\hline
\multicolumn{5}{l}{VISTA (VHS, VIKING)+KIDS, Pan-STARRS, SkyMapper, LSST (Lodieu et al. 2012,}\\
\multicolumn{5}{l}{Beaumont \& Magnier 2010, Keller et al. 2007, Ivezic et al. 2008)}\\
\hline
\end{tabular}
\end{center}
\end{table*}

It is now 50 years since the theoretical prediction of the existence of brown dwarfs (Hayashi \& Nakano 1963, Kumar 1963), but still less than two decades from their observational discovery (Rebolo et al. 1995, Nakajima et al. 1995, Oppenheimer et al. 1995). Throughout the 90s, the observational properties of objects in the substellar regime were still relatively unexplored, but the last decade has seen the discovery of large numbers of brown dwarfs making it possible to constrain several fundamental properties. Our understanding of these objects has now evolved into a consensual idea laid down by Whitworth et al. (2007) that, in general terms, ``\emph{(...) brown dwarfs form like stars (...) on a dynamical timescale and by gravitational instability, with an homogeneous initial elemental composition (...)}'' (Padoan \& Nordlund 2002, Hennebelle \& Chabrier 2008). This is not to say that there is an agreement regarding the exact mechanisms that lead to the formation of these objects. Other plausible formation mechanisms have been proposed, such as gravitational instabilities in disks (Stamatellos \& Whitworth 2009, Basu \& Vorobyov 2012), premature ejection from prestellar cores (Reipurth \& Clarke 2001), or photo-erosion of cores (Kroupa \& Bouvier 2003). Observational evidence is therefore paramount to better understand brown dwarfs and the initial mass function, the distribution of masses with which stars and brown dwarfs form, is one of the fundamental properties of their formation process. In this contribution I highlight a selection of recent observational efforts to characterise the low-mass end of the mass function. Several observational and theoretical reviews on star formation covering the IMF have recently been published by Luhman (2012), Jeffries (2012), Kroupa et al. (2012) and Bastian, Covey \& Meyer (2010). The interested reader should seek those for more thoroughly descriptions and a larger pool of relevant references.

\section{Low-mass IMF studies in the Galaxy}
The initial mass function is defined as the number of stars per unit mass interval (or unit logarithmic mass interval), per cubic parsec. It is commonly parameterised either as a linear or logarithmic power-law, or in the log-normal form. In the following, I will summarise the methodology adopted to study the substellar IMF in the field, open clusters, and star forming regions, alluding to their advantages and pitfalls. I will highlight a selection of results for measurements of the $\alpha$ parameter ($\phi($log~$m$)~$\propto$~$m^{-\alpha+1}$) that is commonly derived in the observations of the substellar IMF. 

\begin{table*}[ht]
\caption{Selection of observed mass functions in open clusters.}
\label{openimf}
\begin{center}
\begin{tabular}{llll}
\hline
\\
Cluster & Age (Myr) &$\alpha$ & References  \\
\hline
\\
Pleiades  		& 125 	& 0.6$\pm$0.11	 			& Moraux et al. 2003, Casewell et al. 2007 \\
		  		& 	 	& 				 			& Lodieu et al. 2007 \\
IC 4665            	& 30 	& $\sim$0.6 					& de Wit et al. 2006, Lodieu et al. 2011 \\
Blanco 1  		& 100  	& 0.69$\pm$0.15			& Moraux et al. 2007 \\
Alpha Per  		& 80  	& 0.59$\pm$0.05			& Barrado y Navascu\'es et al. 2002 \\
		  		& 	  	& 						& Lodieu et al. 2005, Lodieu et al. 2012 \\
IC 2391  		& 50  	& Consistent with Pleiades	& Barrado y Navascu\'s et al. 2004, Boudreault \&  \\
		  		& 	  	& 							& Bailer-Jones 2009, Spezzi et al. 2009 \\
Praesepe		& 600  	& Consistent with Pleiades 	& Boudreault et al. 2010 \\
Hyades		  	& 600 	& Brown dwarfs depleted		& Bouvier et al. 2008 \\
\hline
\end{tabular}
\end{center}
\end{table*}

\subsection{The field}
To date over one thousand of brown dwarfs have been found in the local Galactic field. Their census is done in a two-step approach, with candidate low-temperature objects found through large photometric surveys and extensive spectroscopic follow-ups to measure spectral types and weed-out contaminants. To determine luminosities and derive the space density, parallax measurements are needed but difficult to obtain for large samples, and empirical relations between magnitude and spectral type are used in these cases to estimate distances. The field population contains brown dwarfs of different ages, and it is therefore not possible to use the age-dependent mass-luminosity relation in order to estimate individual masses, and construct the mass function. Instead, statistical simulations of the expected luminosity function are made probing various forms of the mass function, the Galactic birthrate, brown dwarf evolutionary models, and binarity. These are compared to the observed luminosity function derived for a magnitude-- or volume--limited sample corrected for observational bias.  

Some of the advantages of studying the substellar mass function in the field are the fact that there is a larger number of benchmark systems, evolutionary models are more mature for these ages, and reddening is not significant in the solar neighbourhood. However, the mass function cannot be directly measured and its determination is heavily model-dependent. Furthermore, brown dwarfs are fainter when older, compromising the access to the lower mass end of the mass function. Accurate estimates of distances are also difficult to obtain, and the assessment of completeness, contamination, and correction for observational bias introduce additional sources of error.

Table~\ref{fieldimf} shows a collection of some of the recent observational studies of the field mass function, where different ground and space observatories have been used to conduct large surveys. Although the results appear to converge to a field substellar mass function where $\alpha$~$\le$~0, the aforementioned pitfalls of the current methodology limit the accuracy of this result, that is certain to mature in the coming years with the conclusion of on-going surveys also mentioned in Table~\ref{fieldimf}.

\subsection{Open clusters}
From the first discoveries, open clusters have been a preferred site for the study of brown dwarfs. The census of these clusters is made easier from the fact that photometrically the members are located over well defined luminosity ranges, share the same distance, are a coherent kinematic group, and are not significantly affected by strong extinction. Furthermore, the members of a cluster can be assumed to be coeval making several methods available for age determination. It is therefore possible to derive relatively accurate luminosity functions even without having confirmed spectroscopically each member. Beyond photometric colours, other membership indicators are abundant, such as common proper motion, chromospheric activity, remnants of circumstellar disks, lithium depletion, spectral gravity-sensitive features, etc. Once a census of the cluster has been completed, the luminosities derive for each individual member are converted to masses using evolutionary models. 

Table~\ref{openimf} contains a few examples of studies of open clusters. Overall, the agreement of the mass functions of these clusters to that of the Pleiades is remarkable, with an $\alpha$$\sim$0.6. However, the effects of dynamic segregation (e.g., the Hyades) where the lowest mass objects are the first to be ejected from a cluster, or the extent to which unresolved binaries affect the mass estimates, need to be better understood and characterised.   

\begin{table*}
\caption{Examples of observed mass functions in young clusters and star-forming regions.}
\label{youngimf}
\begin{center}
\begin{tabular}{llll}
\hline
\\
Cluster & Age (Myr) &$\alpha$ & References  \\
\hline
\\
IC 348  			& 2 	& 0.7$\pm$0.4				& Alves de Oliveira et al. 2013, Muench et al. 2007 \\
		  		& 	 	& 				 			& Luhman et al. 2003 \\
NGC 1333        & 1	 	& 0.6$\pm$0.1 				& Scholz et al. 2012, Winston et al. 2009 \\
$\sigma$~Ori  	& 3  	& 0.6$\pm$0.2				& Pe\~na Ram\'irez et al. 2012, Caballero et al. 2012 \\
			  	& 	  	& 							& B\'ejar et al. 2011\\
$\lambda$~Ori & 2  	& $\sim$0.3						& Bayo et al. 2011 \\
ONC  			& 1  	& 0.3--0.6	(Deficit of brown dwarfs?)				& Weights et al. 2009 (Da Rio et al. 2011) \\
$\rho$~Oph	& 1  	& 0.7 $\pm$0.3 					& Alves de Oliveira et al. 2012, Muzic el al. 2012 \\
				& 	 	& 						 	& Erickson et al. 2011 \\
Upper Sco		& 5 	& Consistent with other clusters				& Lodieu et al. 2013, Slesnick et al. 2008, \\
				& 	 &  (possible excess of BDs below 30~M$_{Jup}$)		& Lodieu et al. 2008 \\
Cha I			& 3  	& Consistent with other clusters 				& Muzic et al. 2011 \\
 				& 	  	& 	 	 				 	& Luhman et al. 2007 \\
Taurus		  	& 1	 	& Variation of IMF peak 	& Luhman et al. 2004 \\
\hline
\end{tabular}
\end{center}
\end{table*}

\subsection{Young clusters and star forming regions}
With the current instrumentation, nearby young clusters are the only locations where the mass function can be studied down to a few Jupiter masses, since brown dwarfs are brighter at these young ages. They share some of the advantages of observing open clusters, in that members are coeval, the distance to clusters is generally well constrained, and members define one coherent kinematic group. However, variable extinction sometimes over A$_{V}$$\sim$50~mag, means observations of a cluster substellar members must be done in the IR regime. Photometrically, the members of the cluster are not so clearly identified any longer, since extinction reddenes field dwarfs increasing contamination, and the IR colours are not as sensitive to spectral type as in the optical. Spectra of each of the candidate members, usually combined with complementary youth indicators (e.g., accretion, disks, gravity sensitive features, proper-motion), are usually required to construct clean brown dwarf census. The next step is to convert the spectral type to a temperature using empirical relations, and the observed magnitudes to luminosities using the distance to the cluster and empirical relations for the bolometric corrections. The individual masses are then determined from comparing the brown dwarf positions on the H--R diagram to evolutionary models. 

In Table~\ref{youngimf} a collection of IMF determinations for nearby clusters is shown. Although several clusters show a substellar IMF consistent, within the errors, with that derived for the Pleiades, some departures have been reported (e.g., for Taurus, Upper Sco, ONC). There are several limitations of studying the IMF in young clusters. For example, evolutionary models are very uncertain at young ages, the effects of magnetic activity or strong episodic accretion on determination of luminosity are not yet fully understood, and the fact that many of these regions are still forming stars also introduces a degree of uncertainty. In the case of Taurus, the IMF shows clearly a distinct peak in mass than the canonical IMF (e.g., Luhman et al. 2004). For Upper Sco (e.g., Lodieu et al. 2013) and the ONC (e.g., Da Rio et al. 2011) however, the differences of the IMF to other clusters appear at the lower masses of the substellar IMF, where the uncertainties from the limitations mentioned above could be significant. 

\section{Open questions and outlook}
The substellar mass function derived for young and open clusters appears to be approximately consistent with an $\alpha$$\sim$0.6. Although there are clusters showing a departure from such relation, further work needs to be done to understand these results. Besides the method's limitations, the discrepancies in mass determination arising simply from using different evolutionary models and empirical relations at young ages also undermine a direct comparison of results. Taken at face value, the substellar field mass function described by an $\alpha$$\sim$0 could be evidence for a variation to the substellar mass function of open and young clusters. However, the intrinsic differences in methodology together with a large number of assumptions that are adopted in deriving the field mass function, mean that these results are not yet mature enough to draw such conclusion. 

Some of the biggest challenges for advancing the study of the mass function in the substellar regime are to improve the quality of the census covering similar mass ranges down to the very lowest masses, and to develop more accurate evolutionary models, specially at young ages. Several future missions will greatly advance the observations of these objects, for example, Gaia, Euclid, JWST, larger ground telescopes like the ELT or the TMT. They will provide access to the lowest mass regime in the solar neighbourhood but also allow the study of brown dwarfs in distant clusters, extreme environments, and even other galaxies.  
 
\begin{acknowledgements}
I am grateful to the LOC and SOC for an excellent organisation, and also to all the participants that actively contributed to the many timely and lively discussions that have made this conference a memorable event.
\end{acknowledgements}

\bibliographystyle{aa}

\end{document}